\documentclass[%
 reprint,
 amsmath,amssymb,
 aps,
]{revtex4-1}
\usepackage{graphicx}
\usepackage{dcolumn}
\usepackage{bm}

\usepackage{mathrsfs}
\usepackage[colorlinks]{hyperref}
\makeatletter

\newcommand{\Rmnum}[1]{\expandafter\@slowromancap\romannumeral #1@}
\makeatother
\begin{document}

\title{Quantifying entanglement in terms of an operational way}
\author{Deng-hui Yu$^1$}
\author{Chang-shui Yu$^{1,2}$}
\email{ycs@dlut.edu.cn}
\affiliation{$^1$School of Physics and Optoelectronic Technology, Dalian University of
Technology, Dalian 116024, China }

\affiliation{$^2$DUT-BSU Joint Institute, Dalian University of Technology, Dalian 116024, China}

\begin{abstract}
Quantifying entanglement is one of the most important tasks in the
entanglement theory. In this paper, we establish entanglement monotones in
terms of an operational approach, which is closely connected with the state
conversion from pure states to the objective state by the local operations and classical
communications (LOCC).  It is shown that any good entanglement quantifier defined on pure
states can induce an entanglement monotone for all density matrices. We especially show that our entanglement monotone is the maximal one among all that have the same form for pure states. In some particular cases, our proposed entanglement
monotones turned to be equivalent to the convex roof construction, which
hence gains an operational meaning. Some examples are given to demonstrates
the different cases.
\end{abstract}

\maketitle

\section{Introduction}

Entanglement is one of the most intriguing quantum features \cite{p1,p2} and
plays an important role in many quantum information processing tasks \cite{p3,p4}, so quantum entanglement has been recognized as
a key physical resource in quantum information \cite{p5,p6,p7,p8,a,b,b1,b2}%
. Quantification of entanglement, triggering the various researches on the
quantum resource theory \cite{p9,p10,p11,p12,p13,p14,p15,p16,p17,p18,p19,a1,a2,a3,a4,a5}%
, has attracted wide interest for several decades, however, quite limited
progress has been made up to now, due to the good understanding of
entanglement only restricted to bipartite pure states and low-dimensional
mixed states \cite{p20,p21,p22,p23,p24,p25,p26,p27,p28}%
.

The quantification of any quantum resource actually aims to quantitatively
characterize the corresponding quantum feature in a mathematically rigorous
framework \cite{p9}. As to entanglement, a good quantifier should be an
entanglement monotone which is vanishing for separable states and not
increased under local operations and classical communications (LOCC) \cite%
{p27}. There are various such entanglement monotones, however, only a few of
them have the obvious operational meanings. For example, distillable
entanglement quantifies the conversion rate of some standard state
(maximally entangled state) from the given states in the asymptotic regime,
while entanglement cost quantifies the rate of the expected state
asymptotically prepared from some standard state \cite{p27,p29,p30}.
Although the relative entropy of entanglement \cite{p31} defined by the
nearest distance from a given state to the set of separable states based on
the "distance", the relative entropy could have an operational meaning, but
most of the distance-based measure has only the geometric meaning. The
convex roof construction \cite{p32,p33} is a useful approach to
establish an entanglement monotone, which generally has no explicit
operational meaning, but entanglement of formation \cite{p27} can be closely
related to the entanglement cost in the asymptotic regime \cite{p30}.
Similarly, the negativity has not a striking operational meaning \cite{p34},
but the logarithmic negativity provides an upper bound to distillable
entanglement \cite{p35}.  The different ways to quantify entanglement
usually convey different understandings of entanglement, in particular,
their potential operational meanings are  usually connected with different
quantum information processing tasks. How to explore an  operational
approach to quantify entanglement is still an important and significant
topic in the entanglement theory.

In this paper, we proposed an operational way to build entanglement
monotones similar to our previous approach for coherence \cite{p19}. We consider that some pure input states are converted to the common objective
quantum state by LOCC. It is shown that the entanglement of the objective
quantum state can be well characterized by the least entanglement of the
pure input states. We prove that any given pure-state entanglement monotone $%
F$ can induce a good entanglement monotone for a general quantum state, and
especially that our entanglement monotone is the largest one among all the
entanglement monotones that takes the same value for pure states as $F$. We
also show that our entanglement monotone is equivalent to the entanglement
monotone in terms of the convex roof construction, if the convexity is
imposed. As demonstrations, we show that if the chosen pure-state
entanglement monotone linearly depends on the Schmidt coefficients, or we
use the two-qubit concurrence as pure-state measure, our entanglement
monotone will be equal to that of the convex roof construction. In addition,
an analytically computable example indicates that our approach induces an
entirely new entanglement monotone. This paper is organized as follows. In
Sec. II, we directly build the entanglement monotone based on the state
conversion, and then show that our entanglement monotone is the maximal one.
In Sec. III, we study how our method is related to the convex roof
construction. In Sec. IV, we demonstrate several examples in various cases.
The conclusion and discussion is given in Sec. V.

\section{Entanglement monotone based on the state conversion}

Let's consider a bipartite quantum state $\rho =\sum_{i}p_{i}|\psi
_{i}\rangle \left\langle \psi _{i}\right\vert $ with an alternative
pure-state realization $\left\{ p_{i},|\psi _{i}\rangle \right\} $. We use $%
\lambda ^{\downarrow }(|\psi _{i}\rangle )$ denote the Schmidt vector of the
state $|\psi _{i}\rangle $ with the Schmidt coefficients in decreasing
order. It is shown in Ref.~\cite{p36} that if there exists a
bipartite pure state $|\varphi \rangle $ with $\lambda ^{\downarrow
}(|\varphi \rangle )\prec \sum_{i}p_{i}\lambda ^{\downarrow }(|\psi
_{i}\rangle )$ where `$\prec $' is the majorization \cite%
{p37,p38,p39}, one can always find a LOCC to transform the state $%
|\varphi \rangle $ to the state $\rho $. It is obvious that for a fixed
density matrix $\rho $, the state $|\varphi \rangle $ is not unique. In
fact, with the entanglement taken in account, one can also notice that all
these pure states $|\varphi \rangle $ don't necessary have the equal amount
of entanglement. Let $\mathbb{L}(\rho )$ denote the set of pure states which
can be transformed into $\rho $ by LOCC and $F(\cdot )$ denote an arbitary
entanglement monotone of pure states, we are always able to define an
entanglement quantifier for $\rho $ by the smallest amount of entanglement
of the pure states in $\mathbb{L}(\rho )$, which can be given in the
following rigorous way.

\textbf{Theorem 1}.-For any bipartite quantum state $\rho $, let $\mathbb{L}%
(\rho )$ be the set of pure states which can be transformed into $\rho $ by
LOCC, then
\begin{equation}
\mathscr{E}_{F}(\rho )=\mathop{\inf}\limits_{|\varphi \rangle \in \mathbb{L}%
(\rho )}F(|\varphi \rangle )  \label{de}
\end{equation}%
is an entanglement monotone, where the subscript $F$ denotes the chosen
entanglement monotone $F(\cdot )$ of pure states.

\textbf{Proof.} (\textit{Vanishing for separable states}) At first, we'd
like to show that if a state $\varrho $ is separable, there must exist a
separable pure state in the set $\mathbb{L}(\rho )$. To show this, one can
note that any separable state $\varrho $ can be expressed as a convex
combination of some pure product states $\{p_{i},|\phi _{i}\rangle \}$, so $%
\sum_{i}p_{i}\lambda ^{\downarrow }(|\phi _{i}\rangle )$$=(1,0,0,\cdots )$
which majorizes the Schmidt vector $\lambda ^{\downarrow }(\cdot )$ of any
pure product state $|\phi \rangle $. From Ref.~\cite{p36}, it is
easily found that $\varrho $ \ can be converted form a pure product state $%
|\phi \rangle $ by LOCC, which shows $F(\varrho )=0.$

Conversely, if $F(\varrho )=0$, the definition implies there exists pure
product state can be transformed into $\varrho $ by LOCC, thus $\varrho $ is
separable.

(\textit{Monotonicity}) Suppose $\varepsilon $ is an arbitrary LOCC and $%
\sigma =\varepsilon (\rho )$. Let $|\psi \rangle $ be the optimal state in $%
\mathbb{L}(\rho )$ such that $\mathscr{E}_{F}(\rho )=F(|\psi \rangle )$.
Based on the definition of $\mathscr{E}_{F}(\rho )$, we have $|\psi \rangle
\ $can be converted into $\rho$ by LOCC. In addition, $\sigma
=\varepsilon (\rho )$, one can find that $|\psi \rangle $ can also be
converted into $\sigma $ by LOCC, i.e., $|\psi \rangle \in \mathbb{L}(\sigma
)$, which implies $\mathscr{E}_{F}(\rho )=F(|\psi \rangle )\geq \mathscr{E}%
_{F}(\sigma )$.

(\textit{Strong monotonicity}) Suppose $|\psi \rangle $ is the optimal state
in $\mathbb{L}(\rho )$ such that $\mathscr{E}_{F}(\rho )=F(|\psi \rangle )$.
It means that there exists a decomposition $\{t_{i},|\varphi _{i}\rangle \}$
of $\rho $ with
\[
\lambda ^{\downarrow }(|\psi \rangle )\prec \sum_{i}t_{i}\lambda
^{\downarrow }(|\varphi _{i}\rangle ).
\]%
That is, $|\psi \rangle $ can be converted to $\{t_{i},|\varphi _{i}\rangle
\}$. Let an LOCC $\$$ with its Kraus operators $\{M_{k}\}$ performed on the
state $\rho $ with
\begin{align}
p_{k}=& \text{Tr}(M_{k}\rho M_{k}^{\dagger }),  \nonumber \\
\rho _{k}=& M_{k}\rho M_{k}^{\dagger }/p_{k}.  \label{dec2}
\end{align}%
Substituting the decomposition $\{t_{i},|\varphi _{i}\rangle \}$ into Eq. (%
\ref{dec2}), one will obtain
\begin{align}
p_{k}& =\text{Tr}\sum_{i}t_{i}M_{k}|\varphi _{i}\rangle \langle \varphi
_{i}|M_{k}^{\dagger }=\sum_{i}q_{ik}t_{i},  \nonumber \\
\rho _{k}=\sum_{i}t_{i}& M_{k}|\varphi _{i}\rangle \langle \varphi
_{i}|M_{k}^{\dagger }/p_{k}=\sum_{i}\frac{q_{ik}t_{i}}{p_{k}}|\phi
_{ik}\rangle \langle \phi _{ik}|,  \label{rhok}
\end{align}%
with%
\begin{align}
q_{ik}& =\text{Tr}(M_{k}|\varphi _{i}\rangle \langle \varphi
_{i}|M_{k}^{\dagger }),  \nonumber \\
|\phi _{ik}\rangle & =M_{k}|\varphi _{i}\rangle /\sqrt{q_{ik}}.
\end{align}%
Since $|\psi \rangle \xrightarrow{LOCC}\{t_{i},|\varphi _{i}\rangle \}$ and $%
\$$ can convert $\{t_{i},|\varphi _{i}\rangle \}$ to $\{t_{i}q_{ik},|\phi
_{ik}\rangle \}$, we have
\begin{align}
\lambda ^{\downarrow }(|\psi \rangle )\prec & \sum_{ik}t_{i}q_{ik}\lambda
^{\downarrow }(|\phi _{ik}\rangle )  \nonumber \\
=& \sum_{k}p_{k}\sum_{i}\frac{t_{i}q_{ik}}{p_{k}}\lambda ^{\downarrow
}(|\phi _{ik}\rangle )  \nonumber \\
=& \sum_{k}p_{k}\lambda ^{\downarrow }(|\psi _{k}\rangle ),  \label{rep}
\end{align}%
where $|\psi _{k}\rangle $ is defined as a pure state satisfying
\begin{equation}
\lambda ^{\downarrow }(|\psi _{k}\rangle )=\sum_{i}\frac{t_{i}q_{ik}}{p_{k}}%
\lambda ^{\downarrow }(|\phi _{ik}\rangle ).  \label{re}
\end{equation}%
Eq. (\ref{rep}) indicates that $|\psi \rangle $ could be transformed into $%
\{p_{k},|\psi _{k}\rangle \}$ by LOCC, so the entanglement monotone $F(\cdot
)$ gives
\begin{equation}
F(|\psi \rangle )\geq \sum_{k}p_{k}F(|\psi _{k}\rangle ).
\end{equation}%
\ In addition, Eq. (\ref{rhok}) and Eq. (\ref{re}) show $|\psi _{k}\rangle
\in \mathbb{L}(\rho _{k})$, thus
\begin{equation}
F(|\psi _{k}\rangle )\geq \mathscr{E}_{F}(\rho _{k}).
\end{equation}%
Therefore,%
\begin{align}
\mathscr{E}_{F}(\rho )=F(|\psi \rangle )\geq & \sum_{k}p_{k}F(|\psi
_{k}\rangle )  \nonumber \\
\geq & \sum_{k}p_{k}\mathscr{E}_{F}(\rho _{k}),
\end{align}%
which is the strong monotonicity.$\hfill \blacksquare $

One can find that the set $\mathbb{L}(\rho )$ is actually defined by the
state $\vert\psi \rangle $ subject to the majorization relation $\lambda
^{\downarrow }(|\psi \rangle )\prec \sum_{i}p_{i}\lambda ^{\downarrow
}(|\psi _{i}\rangle )$ with $\{p_{i},|\psi _{i}\rangle \}$ denoting the
decomposition of the state $\rho $. However, from the above proofs, an
important relation is
\begin{equation}
\lambda ^{\downarrow }(|\phi \rangle )=\sum_{i}p_{i}\lambda ^{\downarrow
}(|\psi _{i}\rangle ),  \label{rela}
\end{equation}%
where $|\phi \rangle $ is a pure state. It is obvious $\lambda ^{\downarrow
}(|\psi \rangle )\prec \lambda ^{\downarrow }(|\phi \rangle )$ which implies
$F(|\phi \rangle )\leq F(|\psi \rangle )$. Thus the set $\mathbb{L}(\rho )$
in Eq. (\ref{de}) can be replaced by its subset $\mathbb{Q}(\rho )\subset
\mathbb{L}(\rho )$, where $\mathbb{Q}(\rho )$ covers all the pure states $%
|\phi \rangle $ satisfying Eq. (\ref{rela}).

Theorem 1 has provided us with an operational way to define an entanglement
monotone from a pure-state entanglement monotone $F$. That is, the
entanglement of a state $\rho $ quantifies the least entanglement of the
pure states which can be converted into $\rho $. It is obvious that
different $F$ will induce different $\mathscr{E}_{F}$. In fact, there are
many different entanglement monotones which can be reduced to a fixed
entanglement monotone for pure states, which, to some extent, forms the root
of a fundamental requirement of a general entanglement measure: all
entanglement measures should be reduced to the von Neumann entropy of
entanglement for pure states. Next we will show that our proposed
entanglement monotone $\mathscr{E}_{F}$ is the upper bound of all the
entanglement monotones which are identical to $F$ for pure states.

\textbf{Theorem 2}.-Given an entanglement monotone $E(\rho )$ for any
bipartite density matrix $\rho $ such that $E(|\psi \rangle )=\mathscr{E}%
_{F}(|\psi \rangle )$ holds for any bipartite pure state $|\psi \rangle $,
then $\mathscr{E}_{F}(\rho )\geq E(\rho )$.

\textbf{Proof:} Suppose $|\psi _{0}\rangle $ is the optimal state in $%
\mathbb{L}(\rho )$ such that $\mathscr{E}_{F}(\rho )=F(|\psi _{0}\rangle )$,
then we have
\begin{equation}
\mathscr{E}_{F}(\rho )=F(|\psi _{0}\rangle )=E(|\psi _{0}\rangle )\geq
E(\rho ),
\end{equation}%
the last inequality is due to the monotonicity of $E$.$\hfill \blacksquare $

\section{Relation with the convex roof construction}

We have shown that $\mathscr{E}_{F}$ is a valid entanglement monotone, so it
can be safely used to quantify entanglement of a state. However, sometimes
some additional properties are also imposed. One example of the properties
is the concept of convexity. Next we will give the sufficient and necessary
condition for a convex $\mathscr{E}_{F}.$

\textbf{Theorem 3}.-For bipartite $n$-dimensional quantum states, the
following statements are equivalent to each other:

(I) $\mathscr{E}_{F}(\rho )$ is convex.

(II) $\mathscr{E}_{F}(\rho )$ is equivalent to the convex roof construction
in terms of $F\left( \cdot \right) $.

(III) For any $\rho$, the optimal pure state $|\phi _{0}\rangle \in \mathbb{Q%
}(\rho )$ and the related decomposition $\{q_k, |\varphi_k\rangle\}$
satisfy: (1) $F(|\phi _{0}\rangle)=\sum_kq_kF(|\varphi_k\rangle)$, (2) $%
\{q_k, |\varphi_k\rangle\}$ is the optimal decomposition of $\rho $ for the
convex roof construction.

(IV) $F$ satisfies: (1) $F(\cdot )$ should be a linear function of the
decreasing order Schmidt coefficients of a pure state, or (2) for all $n$%
-dimensional states $\rho $, there should be an optimal pure-state
decomposition for the convex roof construction with all the pure states
owing the same Schmidt coefficients.

\textbf{Proof. }Let $|\phi _{0}\rangle \in \mathbb{Q}(\rho )$ be the optimal
pure state for $\mathscr{E}_{F}$, then there exists a decomposition $%
\{q_{k},|\varphi _{k}\rangle \}$ corresponding to $|\phi _{0}\rangle $ such
that Eq. (\ref{rela}) holds. If $\mathscr{E}_{F}$ is convex, we will arrive
at
\begin{equation}
F(|\phi _{0}\rangle )=\mathscr{E}_{F}(\rho )\leq \sum_{k}q_{k}F(|\varphi
_{k}\rangle ).
\end{equation}
A general entanglement monotone $F(\cdot )$ for a bipartite pure state can
always be expressed as a concave function $f$ of the Schmidt coefficients of
the pure state, namely, $f(\lambda(\cdot))=F(\cdot)$ \cite{p33}. From the concavity, we
have $f(\lambda(|\phi _{0}\rangle ))\geq
\sum_{k}q_{k}f(\lambda(|\varphi_{k}\rangle) )$, namely, $F(|\phi _{0}\rangle
)\geq \sum_{k}q_{k}F(|\varphi _{k}\rangle )$. Thus for the optimal state $%
|\phi _{0}\rangle $ and its corresponding decomposition $\{q_{k},|\varphi
_{k}\rangle \}$ of $\rho $, we have
\begin{align}
f(\lambda(|\phi _{0}\rangle))&=\sum_{k}q_{k}f(\lambda(|\varphi _{k}\rangle)
),  \nonumber \\
F(|\phi _{0}\rangle )&=\sum_{k}q_{k}F(|\varphi _{k}\rangle ),  \label{equal}
\end{align}%
which implies the decomposition $\{q_{k},|\varphi _{k}\rangle \}$ achieving $%
\min_{\{p_{i},|\psi _{i}\rangle \}}\sum_{i}p_{i}F(|\psi _{i}\rangle )$ and $%
\mathscr{E}_F$ equal to the minimum. Thus one can arrive at (II) and (III)
from (I). Since Eq. (\ref{equal}) should be satisfied for any $n$%
-dimensional density matrix $\rho $, one can easily find that (1) $F(\cdot )$
should be a linear function of the Schmidt coefficients of a pure state, or
(2) for all $n$-dimensional states $\rho $, there should be an optimal
pure-state decomposition for the convex roof construction with all the pure
states owing the same Schmidt coefficients. Thus we can reach (IV) from (I).

Conversely, if (II) or (III) holds, (I) is clearly holds. If (IV) (1) holds,
then $f(\lambda(|\phi\rangle))=\sum_ip_if(\lambda(|\psi_i\rangle))$ and $%
F(|\phi\rangle)=\sum_ip_iF(|\psi_i\rangle)$ hold for all $|\phi\rangle\in%
\mathbb{Q}(\rho)$ and the related decomposition $\{p_i,|\psi_i\rangle\}$.
Note that $F(|\phi_0\rangle)$ reach the minimum in $\mathbb{Q}(\rho)$, thus
the decomposition $\{q_{k},|\varphi _{k}\rangle \}$ related to $%
|\phi_0\rangle$ achieving the minimum of the convex roof. Thus $\mathscr{E}_F
$ equals to the convex roof and inherits the convexity. If (IV) (2) holds,
suppose the particular decomposition is $\{\tilde{p}_j,|\tilde{\psi}%
_i\rangle\}$, denote $|\tilde{\phi}\rangle$ as the state in $\mathbb{Q}(\rho)
$ related to it, then
\begin{align}
\mathscr{E}_F(\rho)\leq F(|\tilde{\phi}\rangle)=\sum_j\tilde{p}_jF(|\tilde{%
\psi}_i\rangle).
\end{align}
Note that the summation above equals to the convex roof. Combining with
Theorem 2, one can see $\mathscr{E}_F$ equals to the convex roof and
inherits the convexity. The proof is completed.$\hfill \blacksquare $

Theorem 3 shows that the convex $\mathscr{E}_{F}(\rho )$ is equivalent to
the convext roof construction. One should note that if theorem 3 is valid
for all $n$, $\mathscr{E}_{F}(\rho )$ will be the same as the convex roof
construction in the whole state space. In addition, one important thing is
that if the convexity isn't imposed, $\mathscr{E}_{F}$ will be an new
entanglement monotone. In the next section, we will give the examples
subject to different cases.

\section{Examples}

\textit{The same as convex roof with the linear }$\mathit{F(\cdot )}$.-As
the first example, we will demonstrate that $\mathscr{E}_{F}$ will be the
convex roof of $F$ with a proper $F$. To do so, we choose the distillable
entanglement monotone $\left\langle E\right\rangle $ for pure states
proposed in Ref.~\cite{p36} as our entanglement monotone $F$. For a
$d$-dimensional pure state $|\varphi \rangle $, the entanglement monotone is
defined by
\begin{equation}
\left\langle E(|\varphi \rangle )\right\rangle =\sum_{n=2}^{d}\mathscr{P}%
_{n}(|\varphi \rangle )\ln n,  \label{def2}
\end{equation}%
where $\mathscr{P}_{n}(|\varphi \rangle )=n(\lambda _{n}-\lambda _{n+1})$
and $\lambda _{n}$ denotes the Schmidt coefficients. From Ref.~\cite{p36}, one can note that $\left\langle E(|\varphi \rangle
)\right\rangle $ can be rewritten as $\left\langle E(|\varphi \rangle
)\right\rangle =\sum_{l=1}^{d}E_{l}(|\varphi \rangle )z_{l}$ with $%
E_{l}(|\varphi \rangle )=\sum_{m=2}^{l}\lambda _{m}$ and $z_{l}=(l-2)\ln
(l-2)+l\ln l-2(l-1)\ln (l-1)\geq 0$. Therefore, $\left\langle E(|\varphi
\rangle )\right\rangle $ is an entanglement monotone, since $E_{l}(|\varphi
\rangle )$ is an entanglement monotone for all $l$. Thu\textbf{s} one can
establish an entanglement monotone $E_{p}(\rho )$ based on our Theorem 1 as%
\begin{equation}
E_{p}(\rho )=\mathop{\inf}\limits_{|\phi \rangle \in \mathbb{L}(\rho
)}\left\langle E(|\phi \rangle )\right\rangle .  \label{d2}
\end{equation}

Based on the definition of $\left\langle E\right\rangle $ in Eq. (\ref{d2}),
one can find that $\left\langle E\right\rangle $ linearly depends on the
Schmidt coefficients $\lambda _{n},$ which means that Theorem 3 is
satisfied. So our established entanglement monotone $E_{p}(\rho )$ is
equivalent to the convex roof construction in terms of the pure-state
entanglement monotone $\left\langle E(|\varphi \rangle )\right\rangle $.

\textit{The same as convex roof for two-qubit concurrence}.-It has been
shown in Ref.~\cite{p24} that there always exists such an
optimal pure-state decomposition of a bipartite density matrix of qubits
that all the pure states have the same concurrence \cite{p40}, i.e., the
Schmidt coefficients for two-qubit states. Thus, one can easily find that
our $\mathscr{E}_{F}$ for qubit states is equal to the convex roof of
concurrence based on our Theorem 3. In other words, if we select $F$ as
concurrence, $\mathscr{E}_{F}$ is convex in the $(2\otimes 2)$-dimensional
Hilbert space.

\textit{A new entanglement monotone}.-The decomposition similar to bipartite
qubit states doesn't always exist for a high-dimensional system in general
cases, thus one can find that $\mathscr{E}_{F}$ will provide a new
entanglement monotone. To give an explicit demonstration, we will consider
the following analytically computable example, by which one will find that $%
\mathscr{E}_{F}$ is different from the convex roof construction.

\textbf{Theorem 4}.- For a $(3\otimes 3)$-dimensional bipartite density
matrix
\begin{equation}
\sigma =\eta |\varphi _{0}\rangle \langle \varphi _{0}|+(1-\eta )|33\rangle
\langle 33|,
\end{equation}%
where $|\varphi _{0}\rangle =c_{1}|11\rangle +c_{2}|22\rangle $ and $%
|k\rangle $ denotes the computational basis,
\begin{equation}
\mathscr{E}_{F}(\sigma )=F(|\theta \rangle ),
\end{equation}%
with $|\theta \rangle $ denoting the pure state with the Schmidt vector $%
\lambda ^{\downarrow }(|\theta \rangle )=\eta \lambda ^{\downarrow
}(|\varphi _{0}\rangle )+(1-\eta )\lambda ^{\downarrow }(|33\rangle )$ and $%
F $ is an entanglement monotone for pure states.

\textbf{Proof}. Consider any decomposition $\{p_{i},|\psi _{i}\rangle \}$ of
$\sigma $ with $\sigma =\sum_{i}p_{i}|\psi _{i}\rangle \langle \psi _{i}|$,
the Hughston-Jozsa-Wootters (HJW) theorem \cite%
{p20,p41} implies that $|\psi _{i}\rangle $ can
always be written as
\begin{equation}
|\psi _{i}\rangle =x_{i}|11\rangle +y_{i}|22\rangle +z_{i}|33\rangle ,
\end{equation}%
where $x_{i},y_{i},z_{i}$ are the amplitudes with $%
|x_{i}|^{2}+|y_{i}|^{2}+|z_{i}|^{2}=1$. Since $\sigma =\sum_{i}p_{i}|\psi
_{i}\rangle \langle \psi _{i}|$, the correponding elements of the right and
left hand sides with respect to the basis $\{|kk\rangle \}$ should be
consistent with each other, which means
\begin{align}
\sum_{i}p_{i}x_{i}y_{i}^{\ast }& =\eta c_{1}c_{2}^{\ast },  \nonumber \\
\sum_{i}p_{i}|x_{i}|^{2}& =\eta |c_{1}|^{2},  \nonumber \\
\sum_{i}p_{i}|y_{i}|^{2}& =\eta |c_{2}|^{2}.
\end{align}%
The Cauchy-Schwarz inequality implies that $|\sum_{i}p_{i}x_{i}y_{i}^{\ast
}|^{2}=(\sum_{i}p_{i}|x_{i}|^{2})(\sum_{i}p_{i}|y_{i}|^{2})$ holds if and
only if $x_{i}=gy_{i}$ for any $i$. Without loss of generality, we'd like to
suppose $|c_{1}|^{2}\geq |c_{2}|^{2}$, then $|g|^{2}=|c_{1}|^{2}/|c_{2}|^{2}%
\geq 1$ and $|x_{i}|^{2}\geq |y_{i}|^{2}$. Thus $\lambda _{1}^{\downarrow
}(|\psi _{i}\rangle )$ equals to $|x_{i}|^{2}$ or $|z_{i}|^{2}$. Therefore,
\begin{align}
& \sum_{i}p_{i}\lambda _{1}^{\downarrow }(|\psi _{i}\rangle )\leq
\sum_{i}p_{i}(|x_{i}|^{2}+|z_{i}|^{2})  \nonumber \\
=& \eta |c_{1}|^{2}+1-\eta =\lambda _{1}^{\downarrow }(|\theta \rangle ),
\label{major}
\end{align}%
where $|\theta \rangle $ is a state with Schmidt vector $\lambda
^{\downarrow }(|\theta \rangle )=\eta \lambda ^{\downarrow }(|\varphi
_{0}\rangle )+(1-\eta )\lambda ^{\downarrow }(|3\rangle |3\rangle )$. Note
that $\lambda ^{\downarrow }(|\theta \rangle )$ has only two non-zero
elements, thus Eq. (\ref{major}) implies
\begin{equation}
\sum_{i}p_{i}\lambda ^{\downarrow }(|\psi _{i}\rangle )\prec \lambda
^{\downarrow }(|\theta \rangle ).
\end{equation}%
That is, any pure state $|\phi \rangle $ in $\mathbb{Q}(\sigma )$ (with $%
\sum_{i}p_{i}\lambda ^{\downarrow }(|\psi _{i}\rangle )=\lambda ^{\downarrow
}(|\phi \rangle )$) satisfies $\lambda ^{\downarrow }(|\phi \rangle )\prec
\lambda ^{\downarrow }(|\theta \rangle )$. So the monotonicity of $F$ shows $%
F(|\phi \rangle )\geq F(|\theta \rangle ),$ which means $|\theta \rangle $
is the optimal pure state in $\mathbb{Q}(\sigma )$, i.e., $\mathscr{E}%
_{F}(\sigma )=F(|\theta \rangle ).\hfill \blacksquare $

Based on Theorem 3, our entanglement monotone equivalent to the convex roof
construction requires the condition (III). For the state $\sigma $, we have $%
\mathscr{E}_{F}(\sigma )=F(|\theta \rangle )$. However, the optimal pure
state $|\theta \rangle $ should correspond to the optimal decomposition with
the average entanglement given by $\eta F(|\varphi _{0}\rangle )$. It's
obvious that $\eta F(|\varphi _{0}\rangle )=F(|\theta \rangle )$ doesn't
hold for general parameters and $F(\cdot )$. Therefore, one can draw the
conclusion that our approach induces a new entanglement monotone.

\section{Conclusion and discussion}

In summary, we have provided an operational way to define an entanglement
monotone. Since all the bipartite pure states can be converted into their
corresponding mixed/pure objective states by LOCC, we define the
entanglement of the objective state by the least entanglement of the pure
state which can be converted into the objective state of interest. We prove
that any entanglement monotone of pure states can induce an entanglement
monotone of a general quantum state in terms of our approach. In particular,
we prove that our entanglement monotone  is the maximal one among all that
have the same values for pure states as ours. In addition,  we show that if
the convexity is considered, our approach will be equivalent to the convex
roof construction. Thus our approach can provide the operational meaning for
the entanglement monotone based on the convex roof construction.  Finally,
we would like to emphasize that our approach could also be feasible for the
quantification of other quantum resources. This job could motivate the
relevant research on the state conversion by free operations.

\section{Acknowledgements}

This work was supported by the National Natural Science Foundation of China
(Grant No. 11775040, 12011530014 and 11375036) and the Fundamental Research Funds for the
Central Universities (Grant No. DUT20LAB203).


\bibliography{reference}

\end{document}